\documentclass[journal]{IEEEtran}
\ifCLASSINFOpdf
\else
\fi

\usepackage{graphicx}
\usepackage[misc]{ifsym}
\usepackage{url}
\usepackage{hyperref}
\usepackage{amsmath}

\usepackage{caption} 
\usepackage{subfigure}
\usepackage{multirow}
\usepackage{algorithm}
\usepackage{algorithmic}
\begin{document}
%
\title{LearningCC: An online learning approach for congestion control}
%
%
%

\author{Songyang Zhang,
        \thanks{}
\thanks{S. Zhang is with College of Computer Science and
Engineering, Northeastern University, Shenyang, China. (e-mail: sonyang.chang@foxmail.com; leiweimin@ise.neu.edu.cn)}}

\markboth{Journal of \LaTeX\ Class Files,~Vol.~14, No.~8, August~2015}%
{Shell \MakeLowercase{\textit{et al.}}: Bare Demo of IEEEtran.cls for IEEE Journals}
%



\maketitle

\begin{abstract}
Recently, much effort has been devoted by researchers from both academia and industry to develop novel congestion control methods. LearningCC is presented in this letter, in which the congestion control problem is solved by reinforce learning approach. Instead of adjusting the congestion window with fixed policy, there are serval options for an endpoint to choose. To predict the best option is a hard task. Each option is mapped as an arm of a bandit machine. The endpoint can learn to determine the optimal choice through trial and error method. Experiments are performed on ns3 platform to verify the effectiveness of LearningCC by comparing with other benchmark algorithms. Results indicate it can achieve lower transmission delay than loss based algorithms. Especially, we found LearningCC makes significant improvement in link suffering from random loss.
\end{abstract}


%
\IEEEpeerreviewmaketitle

\section{Introduction}
\IEEEPARstart{W}{ithout} congestion control mechanism, it is doubtful whether the Internet can evolve into today's scale, which severs more than half population of the world. By dynamically adjusting delivery rate, congestion control attempts to make efficient utilization of channel resource effectively while preventing overloading the network. After the first congestion collapse even happens in 1986, Jocobson \cite{Jacobson1988Congestion} recommended to apply additive-increase/multiplicative-decrease (AIMD) for rate control rule in TCP, which is later developed into Reno. Since then, congestion control is an active research topic.  

Reno takes packet loss event as congestion signal. Reno flow can send one more packet into network to probe the maximum available bandwidth in every round trip time (rtt). On detecting packet loss event, a connection will reduce the congestion window by half to alleviate congestion. 

Even the AIMD law is said to save the Internet from congestion collapse, it may still provide degenerated performance over some network scenarios. Firstly, a connection may suffer from long latency. In today's wired networks, routers are configured with excessively large buffer, in which the packet loss events seldom happen. Reno flow will keep increasing the number of inflight packets and many packets will be buffered in routers. Secondly, non-congested packet loss event has detrimental effect on throughput of these loss based rate control algorithms, which is common in wireless networks due to interference and signal attenuation. Reducing the congestion window frequently after random loss events will leave the bottleneck operating at speeds considerably below its capacity. Thirdly, the approach to increase congestion window by one over each rtt may be too slow to probe the maximum available bandwidth in long fat pipe.

Due to these drawbacks, different variants have been proposed for some specific networks.  Cubic \cite{Ha2008CUBIC} is proposed for long fat links. Westwood \cite{Grieco2005Mathematical} is for wireless networks. Recently, to design new algorithm which could achieve maximum delivery rate and simultaneously maintain minimal delay becomes a trend. Vivace \cite{Dong2018PCC} and BBR \cite{Cardwell2016BBR} are examples of such effort. The performance of BBR has been evaluated on ns3 \cite{Zhang2019Evaluation}.

There are works optimizing congestion control strategies through machine intelligence. Remy \cite{Winstein2013TCP} uses offline training to find the optimal mapping from observed network states to control actions. Other works \cite{Li2019QTCP} \cite{Xiao2019TCP} use deep reinforce leaning approach. A large number of epochs are needed to train the control parameters in various network scenarios. The policy lookup process may takes orders of magnitude longer compared to hand-crafted methods. These factors hinder these algorithms to be applied in real networks. 

Vivace searches for better action through online optimization. Time is divided into consecutive monitor intervals (MI). Each MI is devoted to test the implication between an action and the performance, which is measured by utility function. 

In this work, we present LearningCC to solve the congestion control problem by multi-armed bandits framework. Instead of adjusting the congestion window with fixed rule, serval options are provided. Each option is taken as an arm of a slot machine. With the help of reward function, tradeoff is made by sender between exploration and exploitation to determine an optimal choice based on live empirical evidence. The effectiveness of the proposed method is evaluated on a small scale network topology. Simulation results indicate that LearningCC can achieve lower transmission delay compared with Reno/Cubic. It can gain higher throughput when coexisting with Reno/Cubic flows compared with Vivace. What's more, LearningCC achieves the best channel utilization in lossy networks than these benchmark algorithms.
\section{Key design on LearningCC}
The congestion window adjustment rule in AIMD is given in Equation \eqref{eq-aimd}. In Reno, $\alpha=1$ and $\beta=0.5$. Such hand-crafted rule make assumption that packet loss is an indication to congestion. When such assumption is violated, halving the congestion window will achieve inferior performance. It is hard to find an always optimal hand-crafted policy in a wide range of real networks. 
\begin{equation}
\label{eq-aimd}
cwnd\leftarrow
\begin{cases}
cwnd+\frac{\alpha}{cwnd}& \text{when an ack arrives}\\
\beta\cdot cwnd& \text{when loss is detected}
\end{cases}
\end{equation}

Hence, the congestion window update rule in LearningCC is updated in a more flexible method. Instead of increasing the congestion window with fixed step in each round, there are several options to update congestion window dynamically in LearningCC. For example, the value on $\alpha$ can be chosen from (1, 2, 3,$\cdots$). For a short time, the congestion window can be increased by 2 ($\alpha=2$) in each rtt to gain better channel utilization. And it may be also feasible to set $\alpha=3$. Since it is hard to know which option can generate best benefit, MAB is used to learn the best action under such uncertain environment.

In MAB, a gambler makes decision on which arm to pull in a K-slot machine. Reward is only observed when an arm is selected and the goal of the gambler is maximize the cumulative reward. Due to lack of oracle perspective, it is hard for gambler to pull the arm always generating the highest reward in each time. The gambler has to pull multiple arms to identify the optimal choice. The gambler will try alternatives to acquire reward distribution information in exploration phase. As time goes by, the gambler gains the information on the reward distribution of each arm. The gambler can exploit the arm that gives the highest reward as much as possible.

During the persistent of a session, the throughput is calculated from acknowledged packets. When a new packet is sent out, the packet state information (bytes, sent\_time, delivered) is recorded in sent\_packet. Here,  $delivered$ counts the total bytes of packets that have been successfully delivered to the peer. When an acknowledged packet arrives, the duration ($\Delta t$) and the delivered packets ($\Delta delivered$) in this round are known. A measurement on throughput can be calculated as Equation \eqref{eq:delivered-rate}.
\begin{equation}
\label{eq:delivered-rate}
rate(n)=\frac{\Delta delivered}{\Delta t}
\end{equation}

In recent proposed congestion control solutions, the goal is to maximize throughput while simultaneously minimizing transmission delay.  The value of $\frac{throughput}{delay}$ is defined as reward.  When an ack arrives, the instant reward of a congestion window update action is defined in Equation \eqref{eq:reward}. The smoothed round trip time ($srtt$) is got by a low pass filter in Equation \eqref{eq-srtt} and $\gamma$ is empirically set as 0.125. When the measurement on rtt is first got, $srtt\gets rtt$.
\begin{equation}
\label{eq:reward}
reward(n)=\frac{rate(n)}{srtt(n)}
\end{equation}
\begin{equation}
\label{eq-srtt}
srtt(n)=(1-\gamma)\cdot srtt(n-1)+\gamma\cdot rtt(n)
\end{equation}

When a new action is chosen, its impact on a netowork system is delayed by one round. The reward can be calculated when theese packets sent after the selection of an action get acknowledged from peer. Each arrival of an ack will generate an instant reward. The exponential filter is applied again to update the reward in Equation \eqref{eq-average-reward} and the factor $\delta$ (0.85) gives more importance to recent instant reward. Here, $i$ is the index of an action.
\begin{equation}
\label{eq-average-reward}
Reward(n,i)=(1-\delta)\cdot Reward(n-1,i)+\delta\cdot reward(n, i)
\end{equation}

In traditional MAB problem, the gambler can choose an arm at each time step. For congestion control, the reward is not instantly observable but is delayed by at least one round after an action is selected. The decision-making process is not triggered by fixed time step but is triggered by congestion event. The congestion event is inferred from increased delay and lost packet. 
\begin{equation}
\label{eq-delay-th}
rtt_{th}=rtt_{base}+\theta\cdot (srtt_{max}-rtt_{base})
\end{equation}

In LearningCC, the values of $rtt_{min}$, $srtt_{max}$ and $rtt_{base}$ are monitored. $rtt_{min}$ is the minimal rtt and $srtt_{max}$ is the maximum smoothed round trip delay during the observation time window. $rtt_{base}$ is the minimal rtt after an action is chosen. In addition to packet loss event, the link is deemed falling into congestion when a new sample rtt exceeds $rtt_{th}$ defined in Equation \eqref{eq-delay-th} and to choose a new congestion window update rule is triggered. $\theta$ is empirically set as 0.8. Firstly, endpoint enters the recover state and the congestion window will be reset as $\beta_{l}\cdot Bw\cdot rtt_{min}$ as shown in Algorithm \ref{alg:backoff}. $\beta_{l}$ is 0.9. The operation to actively reduce congestion window is to alleviate link congestion. $Bw$ is the maximum estimated throughput in 10 rtts and the throughput is got by Equation \eqref{eq:delivered-rate}. Since the endpoint has already responded link congestion event, $rtt_{base}$ will be resampled. The detail on action selection for $\alpha$ is given in Algorithm \ref{alg:acked}. The $\epsilon$-greedy method is applied for decision making. When the random generated value is smaller than $\epsilon$ (0.3), the endpoint enters the exploration procedure (line 7) and an action is randomly chosen from action table ($kActionTable=[1,2,3,4,5]$). Otherwise, the endpoint will choose the action that has maximum reward in reward table during exploitation procedure (line 9). MSS denotes maximum segment size.
\begin{algorithm}[htb] 
\caption{CongestionWindowBackoff} 
\label{alg:backoff} 
\begin{algorithmic}[1]
\REQUIRE ~~\\ 
packet\_number, has\_loss, rtt
\IF{$rtt > rtt_{th}\ \textbf{or}\ has\_loss$}
\IF{$last\_cutback\_!=0\ \textbf{and}\ packet\_number \leq last\_cutback\_$}
\STATE{\textbf{return}}
\ENDIF
\STATE{$cwnd\_\gets \beta_{l}\cdot Bw\cdot rtt_{min}$}
\STATE{$last\_cutback\_\gets largest\_sent$}
\STATE{$acked\_count\_\gets 0$}
\STATE{$action\_chosen\_\gets 0$}
\ENDIF
\end{algorithmic}
\end{algorithm}
\begin{algorithm}[htb] 
\caption{OnPacketAcked} 
\label{alg:acked} 
\begin{algorithmic}[1]
\REQUIRE ~~\\ 
packet\_number,rtt
\IF{$last\_cutback\_!=0\ \textbf{and}\ packet\_number \leq last\_cutback\_$}
\STATE{\textbf{return}}
\ENDIF
\IF{$!action\_chosen\_$}
\STATE{$rtt_{base}\gets \infty$}
\IF{$random.Real(0,1)<\epsilon$}
\STATE{\text{Exploration()}}
\ELSE
\STATE{\text{Exploitation()}}
\STATE{$action\_chosen\_\gets 1$}
\ENDIF
\ENDIF
\IF{$rtt<rtt_{base}$}
\STATE{$rtt_{base}\gets rtt$}
\ENDIF
\STATE{$acked\_count\_++$}
\STATE{$action\gets kActionTable[action\_index\_]$}
\IF{$acked\_count\_*action\geq \frac{cwnd\_}{MSS}$}
\STATE{$cwnd\_+=MSS$}
\STATE{$acked\_count\_\gets 0$}
\ENDIF
\end{algorithmic}
\end{algorithm}
\section{Fluid model of LearningCC}
Following the method in \cite{Grieco2005Mathematical}, the theory throughput of LearningCC is analyzed by fluid model.  $p$ denotes the probability of congestion event. For every acknowledged packet, the increment of $cwnd$ is $\frac{\alpha}{cwnd}$. For every congestion event, the decrement on $cwnd$ is  denoted by $D$. The expected increment on $cwnd$ per update step is then: $\Delta w=(1-p)\frac{\alpha}{cwnd}+pD$. The delivery rate at time t is $x(t)=\frac{cwnd}{rtt}$. The time between update steps is the inter arrival time of two adjacent acks: $\Delta t=\frac{rtt}{cwnd}$. The derivative on $x$ is given in Equation \eqref{eq-rate}.
\begin{equation}
\label{eq-rate}
\dot x=\frac{\mathrm{d}x}{\mathrm{d}t}=\frac{1}{rtt}\frac{\mathrm dw}{\mathrm dt}=\frac{1}{rtt}\frac{\Delta w}{\Delta t}=\frac{x}{rtt}\{(1-p)\frac{\alpha}{cwnd}-pD\}
\end{equation}

For Reno flow, $\alpha=1$, $D=\beta\cdot cwnd$. By substituting $x=\frac{cwnd}{rtt}$, we could further get:
\begin{equation}
\label{eq:aimd-rate}
\dot x=\frac{1-p}{rtt^2}-\beta\cdot x^2\cdot p
\end{equation}

Let $\dot x=0$, the throughput gained by a Reno flow at equilibrium is: 
\begin{equation}
\label{eq:reno-rate}
x_{reno}=\frac{1}{rtt}\sqrt{\frac{(1-p)}{\beta\cdot p}}
\end{equation}

For LearningCC, the increase factor is updated in an online learning fashion. We assum the average increase factor is $\overline\alpha$. $D_l=cwnd-\beta_{l}\cdot Bw\cdot rtt_{min}$. According to Equation \eqref{eq-rate},  the fluid model of a LearningCC flow can be got:
\begin{equation}
\label{eq-reno-rate}
\dot x=\frac{\overline\alpha\cdot (1-p)}{rtt^2}+p\cdot\beta_{l}\cdot x\cdot Bw\cdot\frac{rtt_{min}}{rtt}-p\cdot x^2
\end{equation}

Let $\dot x=0$ and $Bw=x$, the throughput gained by a LearningCC flow at equilibrium is: 
\begin{equation}
\label{eq:reno-rate}
x_{learningCC}=\frac{1}{\sqrt{rtt(rtt-\beta_{l}\cdot rtt_{min})}}\cdot\sqrt{\frac{\overline\alpha\cdot(1-p)}{p}}
\end{equation}

When $\overline\alpha>\frac{1}{\beta}(1-\beta_{l}\frac{rtt_{min}}{rtt})$, $x_{leanringCC}>x_{reno}$. We analyze a simple case. When $rtt=2\cdot rtt_{min}$, the value of $\overline\alpha$ is larger than 1.1 and LearningCC can achieve higher throughput than Reno flow when they traverse the same path. Such requirement seems easy to be met. And we will show by experiments that learningCC is more robust in lossy networks.
\section{Evaluation}
All experiments are running on ns3.26. The code and all scripts to reproduce the results presented in this work can be found in repository \cite{code-learningcc}.

\begin{figure}
\includegraphics[height=1.2in, width=3in]{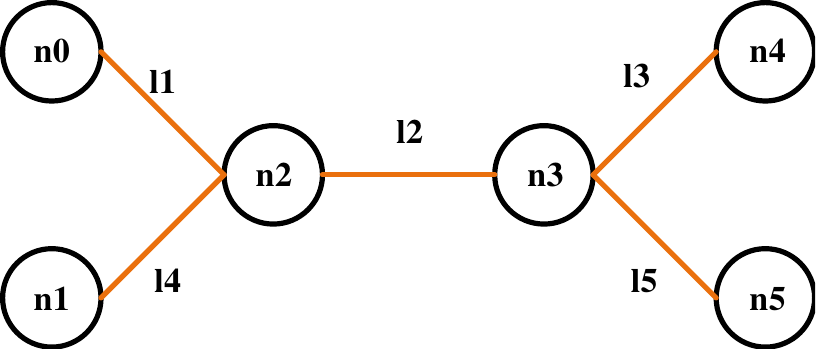}
\caption{Network topology}
\label{Fig:topology}
\end{figure}

The dumbbell topology in Figure \ref{Fig:topology} is built to evaluate LearningCC. These parameters on each link in Table \ref{tab:config} are bandwidth (B, in unit of Mbps), one way propagation delay (D, in unit of milliseconds) and maximum queue delay (Qdelay, in unit of milliseconds). The maximum queue delay is converted to maximum buffer length ($q=BQdelay$) in routers. Four flows are involved. Flow1 and flow2 send packets over pasth1 (n0 to n4). Flow3 and flow4 use path2 (n1 to n5). Each experiment lasts 300 seconds.
\begin{table}[]
\centering
\caption{Configuration on the dumbbell topology}
\label{tab:config}
\scalebox{0.9}{
\begin{tabular}{|c|c|c|c|c|c|}
\hline
\multirow{2}{*}{Case} & l1           & l2         & l3           & l4           & l5           \\ \cline{2-6} 
                      & \multicolumn{5}{c|}{(BW,OWD,Qdelay)}                                   \\ \hline
1                     & (100,10,60)  & (5,10,60)  & (100,10,60)  & (100,10,60)  & (100,10,60)  \\ \hline
2                     & (100,10,120) & (5,10,120) & (100,10,120) & (100,20,120) & (100,10,120) \\ \hline
3                     & (100,30,100) & (6,10,100) & (100,10,100) & (100,20,100) & (100,20,100) \\ \hline
4                     & (100,10,150) & (6,10,150) & (100,10,150) & (100,20,150) & (100,20,150) \\ \hline
5                     & (100,5,90)   & (8,10,90)  & (100,5,90)   & (100,15,90)  & (100,5,90)   \\ \hline
6                     & (100,20,120) & (8,20,120) & (100,20,120) & (100,20,120) & (100,20,120) \\ \hline
\end{tabular}}
\end{table}
\begin{figure}
\centering
\includegraphics[width=3in]{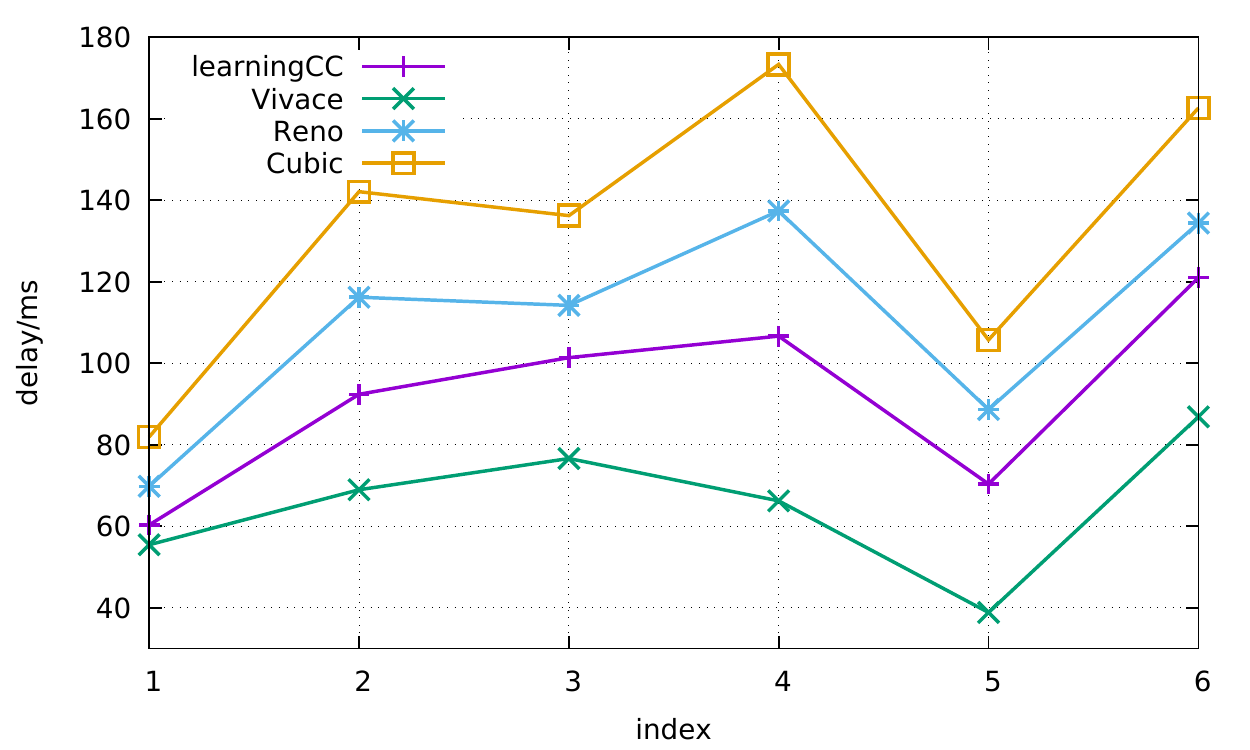}
\caption{Average one way delay}
\label{Fig:owd}
\end{figure}
\begin{figure}
\centering
\includegraphics[width=3in]{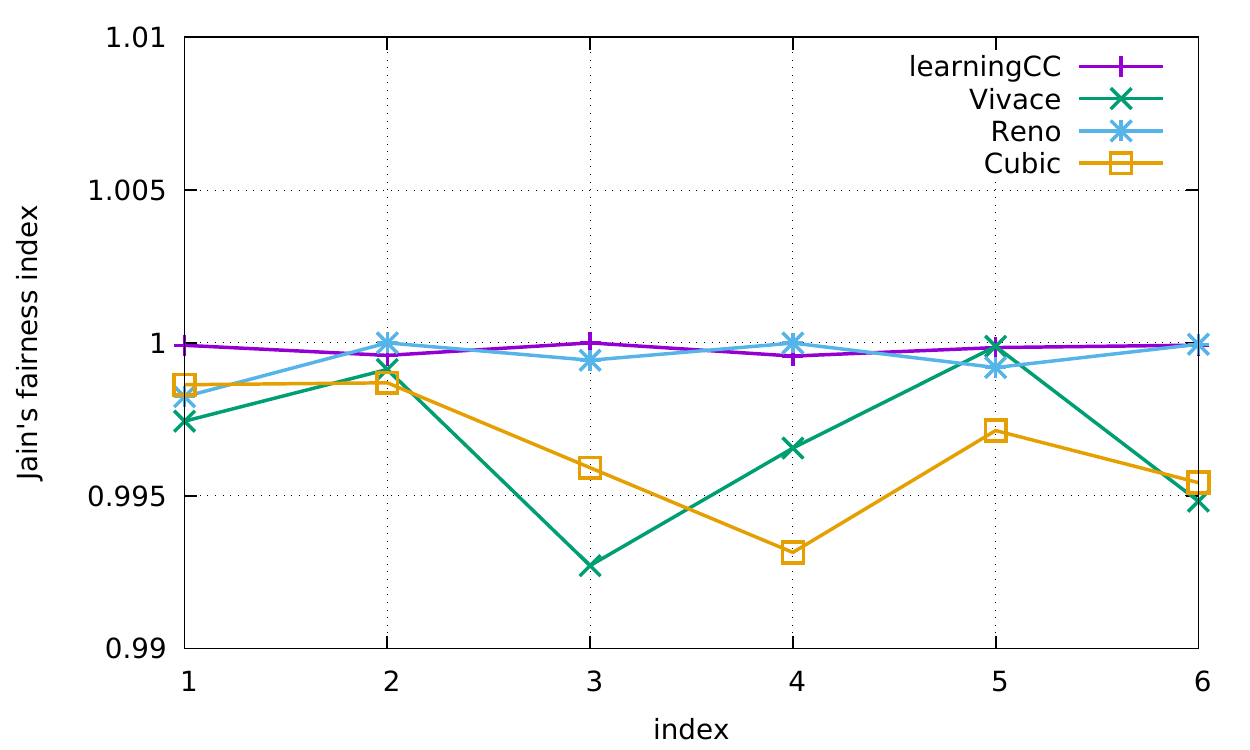}
\caption{Jain's fairness index}
\label{Fig:fairness}
\end{figure}
\begin{figure}
\centering
\includegraphics[width=3in]{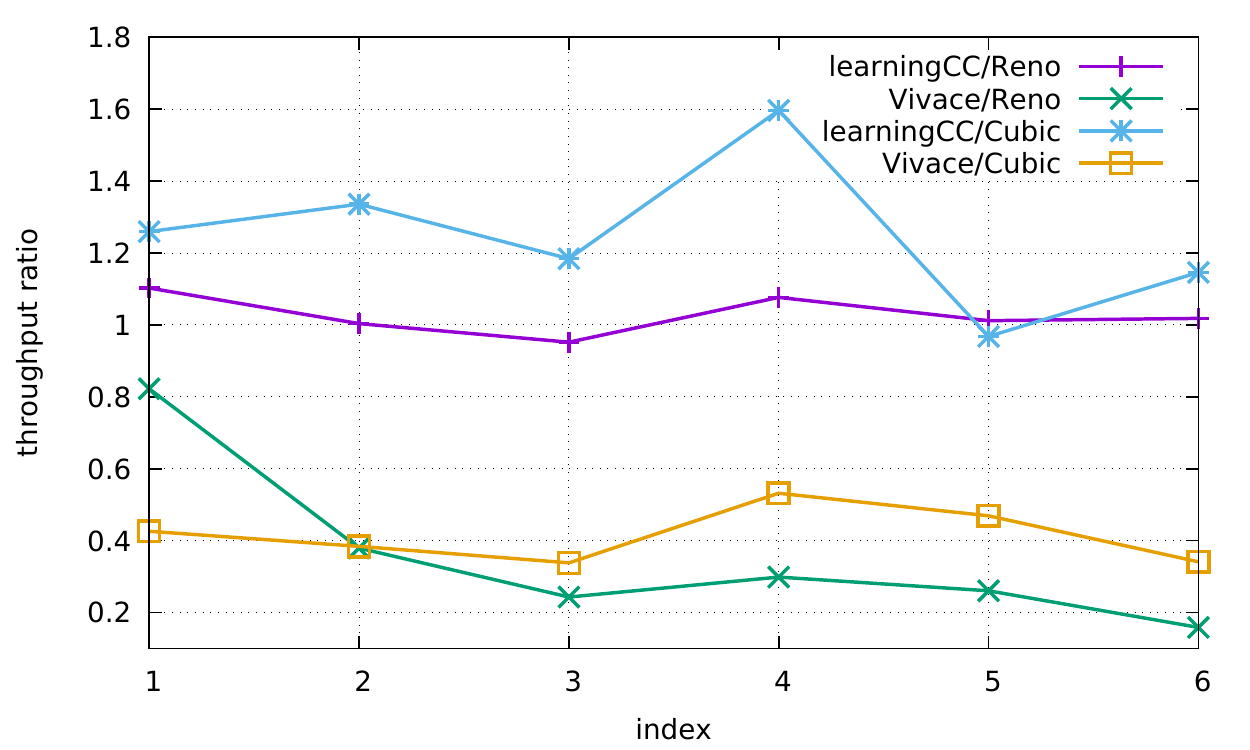}
\caption{Throughput ratio}
\label{Fig:ratio-reno-cubic}
\end{figure}
\begin{figure}
\centering
\includegraphics[width=3in]{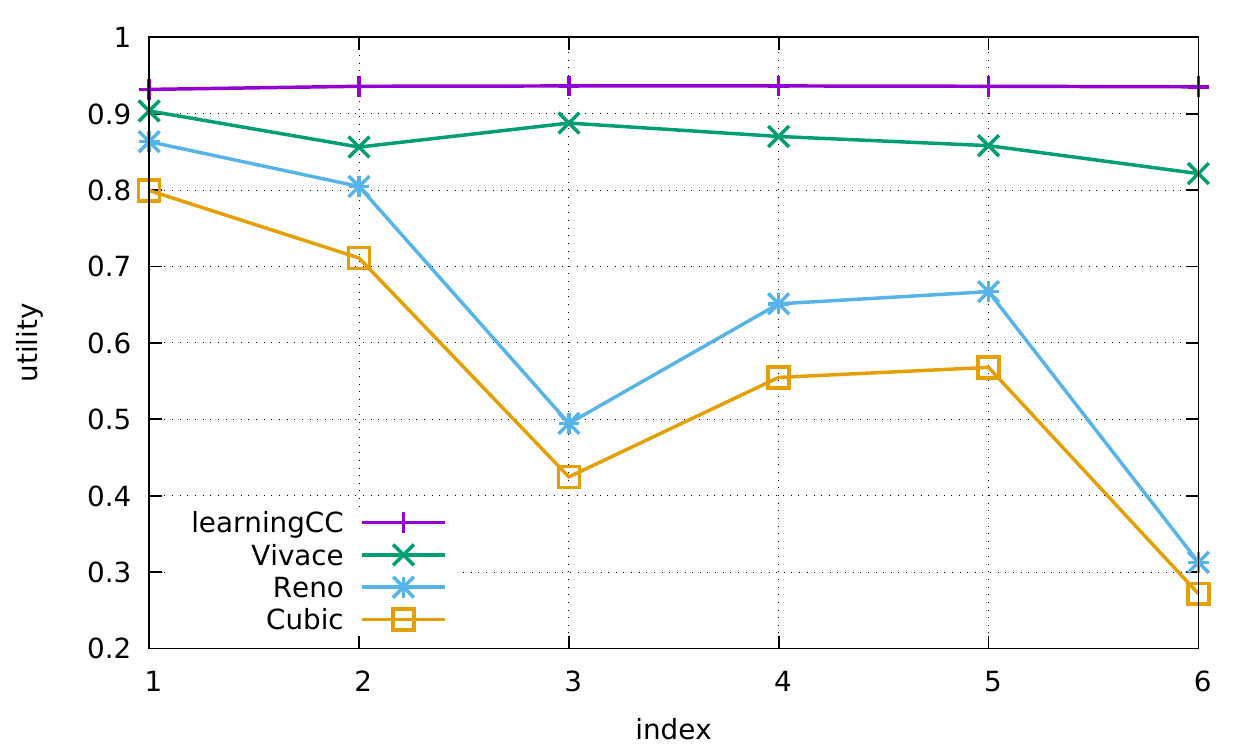}
\caption{Channel Utilization (3.5\%)}
\label{Fig:l35}
\end{figure}
\begin{figure}
\centering
\includegraphics[width=3in]{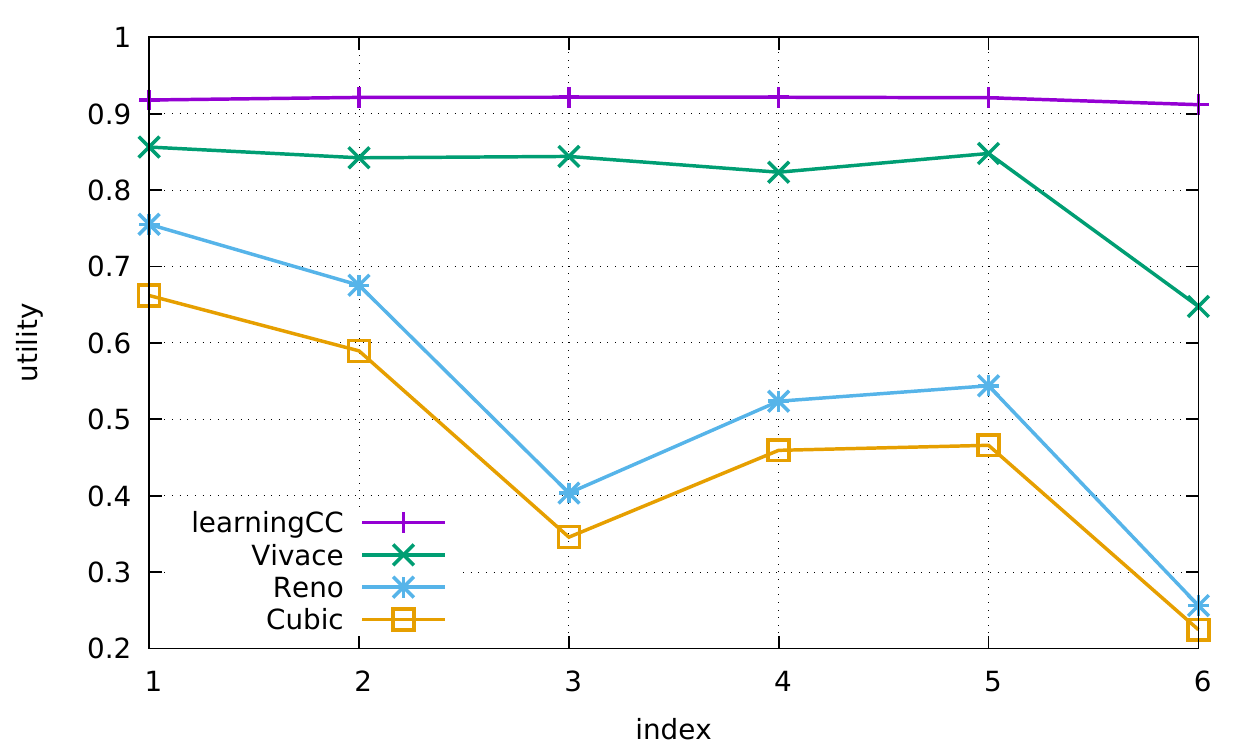}
\caption{Channel Utilization (5\%)}
\label{Fig:l50}
\end{figure}
Protocol fairness is an important indication to reflect whether a flow can converge to a fair bandwidth line when sharing link with other flows with same protocol. The four flows are configured with the same congestion control algorithm (Reno, Cubic, Vivace, or LeraningCC) . Vivace is the closest relevant congestion control mechanism to our design, which also takes an online learning optimization approach.

Once a packet is injected into network, the sent timestamp is tagged into the packet object in ns3. When it arrives to destination, one way transmission delay and its length are recorded. One way transmission delay reflects the buffer occupation status in routers.
\begin{equation}
\label{eq:jain}
J=\frac{(\sum_i^{n}{x_i})^2}{n\cdot(\sum_i^{n}{x_i^2})}
\end{equation}
\begin{equation}
\label{eq:rate}
x_i=\frac{bytes_i}{duration}
\end{equation}

According to collected data, the average one way transmission delay of the two flows on path1 is calculated. The result on delay is given in Figure \ref{Fig:owd}. The Jain's fairness index \cite{Jain1984Quantitative} in Equation \eqref{eq:jain} is exploited to indicate how fair the bandwidth is shared. The closer Jain's fairness index is to 1, the better in terms of bandwidth allocation fairness. $x$ is the average throughput of a session, which is defined in Equation \eqref{eq:rate}. $duration$ is the persistence time of a session and $bytes$ is the length of all received packets. The result on fairness in given in Figure \ref{Fig:fairness}.

Some conclusions can be summarized from these metrics. Compared with Reno and Cubic, Learning CC achieves lower transmission delay. When the delay signal exceeds the predefined threshold, the switch to a different action according to reward values will be triggered. Vivace flows have the lowest transmission delay. In term of fairness, LearningCC and Reno gain better performance.
\subsection{Bandwidth competence}
\begin{equation}
\label{eq:ratio}
ratio=\frac{x_{1}}{x_{2}}
\end{equation}

A route can be shared by many flows with different congestion control algorithms in real networks. The loss based congestion control algorithms still dominate current Internet. It is important for a flow with newly designed algorithm  to achieve better throughput or to avoid being starved by other flows. Such property can motivate the deployment of a new algorithm. In this part, the performance of LearningCC and Vivace is tested when they share route with Reno/Cubic flows. In each test, flow 1 and flow3 take a learning based approach (LearningCC or Vivace) for rate control while Flow2 and flow4 is configured with a loss based method (Reno or Cubic). The throughput ratio is defined in \eqref{eq:ratio} to measure the bandwidth competence ability. $x1$ is the average throughput of flow1 and $x2$ is the average throughput of flow2. The configuration on each link remains the same as the previous part.

The result on throughput ratio is given in Figure \ref{Fig:ratio-reno-cubic}. The throughput of a LearningCC flow is slightly higher than a Reno flow. It means LearningCC can maintain well inter-protocol fairness. LearningCC can gain higher bandwidth when sharing bottleneck with Cubic. But the throughput of a Vivace flow is quite low and Vivace flow is nearly starved by Reno or Cubic flow in each test.
\subsection{Performance in lossy links}
The random packet loss event are common in wireless networks due to interference and signal attenuation. The bottleneck l2 is configured with different random loss rate (1\%,1.5\%, 2\%, 2.5\%, 3\%,3.5\%, 4\%, 4.5\%, 5\%). These values are based on the measurement in wireless network from Uber \cite{quic-uber}. The four flows will take the same algorithm for  rate control. The channel utilization defined in Equation \eqref{eq:util} is computed. $C$ denotes the capacity of bottleneck l2 and $T$ is the running time in each test.
\begin{equation}
\label{eq:util}
util=\frac{\sum_{i}^{n}{bytes_i}}{C\cdot T}
\end{equation}

Due to space limitation, the results on channel utilization of these algorithms when the bottleneck is configured with 3.5\% and 5\% random loss rate are given in Figure \ref{Fig:l35} and Figure \ref{Fig:l50}. The random packet loss event is wrongly interpreting as congestion signal.  Reno/Cubic flows will frequently reduce the congestion window in random loss environment. As shown in Figure \ref{Fig:l35}, Reno flows achieve 31\% channel utilization and Cubic flows achieve 27\% channel utilization in test 6. In all tests, LearningCC makes better channel utilization than Vivace. LearningCC flows achieve channel utilization above 90\% even when the bottleneck introduces 5\% random loss.
\section{Conclusion}
In this letter, a new perspective is provided to solve the congestion control problem with an online learning approach. By mapping each congestion window increment rule to an arm in multi-armed bandit scenario, $\epsilon$-greedy is applied to discovery which decision generates the maximum reward through trial and error. LearningCC is further implemented on ns3 simulator and its performance is compared with three benchmark algorithms. When the small scale network is occupied by LearningCC flows, LearningCC achieves lower transmission delay at the cost of reduced channel utilization. Even Vivace has the lowest transmission delay, Vivace flow maintains lower throughput and nearly gets starved. LearningCC can achieve similar throughput when competing bandwidth with Reno and it maintains higher throughput when sharing bottleneck with Cubic. Most importantly, the channel utilization of LearningCC is less affected by random loss, which makes it fit to be applied in wireless networks.
\ifCLASSOPTIONcaptionsoff
  \newpage
\fi




\begin{thebibliography}{1}
\bibitem{Jacobson1988Congestion}
V.~Jacobson, Congestion avoidance and control, SIGCOMM Comput. Commun. Rev.
  18~(4) (1988) 314--329.

\bibitem{Ha2008CUBIC}
S.~Ha, I.~Rhee, L.~Xu, Cubic: a new tcp-friendly high-speed tcp variant, ACM
  SIGOPS operating systems review 42~(5) (2008) 64--74.

\bibitem{Grieco2005Mathematical}
L.~A. {Grieco}, S.~{Mascolo}, Mathematical analysis of westwood+tcp congestion
  control, IEE Proceedings - Control Theory and Applications 152~(1) (2005)
  35--42.

\bibitem{Dong2018PCC}
M.~Dong, T.~Meng, D.~Zarchy, E.~Arslan, Y.~Gilad, P.~B. Godfrey, M.~Schapira,
  Pcc vivace: Online-learning congestion control, in: Proceedings of the 15th
  USENIX Conference on Networked Systems Design and Implementation, NSDI'18,
  USA, 2018, pp. 343--356.

\bibitem{Cardwell2016BBR}
N.~Cardwell, Y.~Cheng, C.~S. Gunn, S.~H. Yeganeh, V.~Jacobson, Bbr:
  Congestion-based congestion control, Queue 14~(5) (2016) 50:20--50:53.

\bibitem{Zhang2019Evaluation}
S.~Zhang, An evaluation of bbr and its variants (2019).
\newblock \href {http://arxiv.org/abs/arXiv:1909.03673}
  {\path{arXiv:arXiv:1909.03673}}.

\bibitem{Winstein2013TCP}
K.~Winstein, H.~Balakrishnan, Tcp ex machina: Computer-generated congestion
  control, SIGCOMM Comput. Commun. Rev. 43~(4) (2013) 123--134.

\bibitem{Li2019QTCP}
W.~{Li}, F.~{Zhou}, K.~R. {Chowdhury}, W.~{Meleis}, Qtcp: Adaptive congestion
  control with reinforcement learning, IEEE Transactions on Network Science and
  Engineering 6~(3) (2019) 445--458.

\bibitem{Xiao2019TCP}
K.~{Xiao}, S.~{Mao}, J.~K. {Tugnait}, Tcp-drinc: Smart congestion control based
  on deep reinforcement learning, IEEE Access 7 (2019) 11892--11904.

\bibitem{code-learningcc}
Implementation on learningcc,
  \url{https://github.com/SoonyangZhang/learningcc}.

\bibitem{Jain1984Quantitative}
R.~K. Jain, D.-M.~W. Chiu, W.~R. Hawe, A quantitative measure of fairness and
  discrimination for resource allocation in shared computer systems, Eastern
  Research Laboratory, Digital Equipment Corporation, Hudson, MA.

\bibitem{quic-uber}
Employing quic protocol to optimize uber's app performance,
  \url{https://eng.uber.com/employing-quic-protocol/}.

\end{thebibliography}
%

\end{document}